\documentclass{INTERSPEECH2023}
\interspeechcameraready

\usepackage{multirow}

\title{Debiased Automatic Speech Recognition for Dysarthric Speech\\ via Sample Reweighting with Sample Affinity Test}

\name{Eungbeom Kim$^{1\star}$, Yunkee Chae$^{1\star}$, Jaeheon Sim$^1$, Kyogu Lee$^{1, 2}$\thanks{$^\star$ Equal contribution. This work was partly supported by Institute of Information
\& communications Technology Planning \& Evaluation (IITP)
grant funded by the Korea government(MSIT) [No. 2022-0-
00641, XVoice: Multi-Modal Voice Meta Learning, 90\%]
and [NO.2021-0-01343, Artificial Intelligence Graduate School
Program (Seoul National University), 10\%].}}

\address{
  $^1$IPAI, $^2$AIIS, Seoul National University, Seoul, Republic of Korea
  }
\email{\{eb.kim, yunkimo95, sjhoney0112, kglee\}@snu.ac.kr}

\begin{document}

\maketitle

\begin{abstract}
Automatic speech recognition systems based on deep learning are mainly trained under empirical risk minimization (ERM). 
Since ERM utilizes the averaged performance on the data samples regardless of a group such as healthy or dysarthric speakers, ASR systems are unaware of the performance disparities across the groups. 
This results in biased ASR systems whose performance differences among groups are severe. 
In this study, we aim to improve the ASR system in terms of group robustness for dysarthric speakers. 
To achieve our goal, we present a novel approach, sample reweighting with sample affinity test (Re-SAT).
Re-SAT systematically measures the debiasing helpfulness of the given data sample and then mitigates the bias by debiasing helpfulness-based sample reweighting. 
Experimental results demonstrate that Re-SAT contributes to improved ASR performance on dysarthric speech without performance degradation on healthy speech.

\end{abstract}
\noindent\textbf{Index Terms}: speech recognition, debiasing, dysarthric speech

\section{Introduction}
Automatic speech recognition (ASR) performance across groups should be fair regardless of disorder, race, age, and dialect for a trustworthy and inclusive system. 
Although ASR has been improved along with the success of deep learning, ASR systems based on deep learning tend to be vulnerable to biases~\cite{martin2022bias, feng2021quantifying}. 
Consequently, the bias in ASR systems interferes with the ASR system's trustworthiness by causing poor worst group performance.
Applying ASR on dysarthria, which is a type of motor speech disorder, also suffers from the lack of group robustness between healthy and dysarthric speakers. 
In other words, a dysarthric speaker group has a lower performance than a healthy speaker group on ASR. 
This is because dysarthric speech has low intelligibility, whereas deep learning has a tendency to be easily fitted to shortcuts~\cite{beery2018recognition, asgari2022masktune}.
However, most of the previous dysarthric speech recognition studies mainly focus on the sole performance of the ASR system on dysarthric speech. 
In this study, we aim to improve the performance of ASR on dysarthric speech from a debiasing point of view. 
To the best of our knowledge, this is the first work to endeavor to make a debiased ASR system for dysarthric speech.

In this regime, we focus on a debiasing method based on sample reweighting. Recently, sample reweighting such as \cite{nam2020learning, liu2021just} is proposed as a promising solution to handle the bias problem.
Sample reweighting methods mainly consist of 1) estimating the helpfulness of a sample for debiasing and 2) reweighting the sample based on helpfulness.
To estimate the debiasing helpfulness, \cite{nam2020learning} defines \textit{bias-conflicting sample} and \textit{bias-aligned sample} which denote an incorrectly classified sample and a correctly classified sample from an unintended decision rule of a biased model, respectively.
The strategies utilizing those definitions are based on the assumption that upweighting the data samples with a large loss from the biased model, i.e. bias-conflicting samples, leads to improved generalization performance of underfitted groups. 
This assumption, however, hardly holds in real-world problems.
For example, although an outlier might have a large loss value, it is an undesirable strategy to upweight the outlier.

Therefore, we define a new taxonomy to directly categorize the samples by debiasing helpfulness; \textit{bias-blocking sample} denotes a sample that mitigates the model's bias, and \textit{bias-accelerating sample} denotes a sample that accelerates the model's bias when the model is trained on each sample.
To estimate whether a given sample is \textit{bias-blocking sample}, we propose a sample affinity test (SAT) which systematically measures debiasing helpfulness. SAT is based on the novel metric dubbed sample affinity, which denotes a training effect of the given sample to the other samples' loss.
This is inspired by task affinity~\cite{fifty2021efficiently} which estimates inter-task affinity using the loss shift of each task after lookahead training of a given task.
Intuitively, SAT measures debiasing helpfulness using sample affinity of the given sample on the bias-conflicting sample set, based on a one-step lookahead training of a given sample. Unlike the loss-based methods, SAT can filter unhelpful bias-accelerating samples even if the samples are bias-conflicting.

Based on SAT, we propose a novel sample reweighting method, Re-SAT, to challenge the debiasing problem of dysarthric speech recognition. 
Re-SAT consists of four sequential components for implementation. 
The first component estimates bias-conflicting samples based on loss, following \cite{liu2021just}. 
Secondly, the SAT component is activated based on the estimated bias-conflicting samples to accurately measure the debiasing effect of each sample. 
In the third component, the SAT result is normalized through sorting and mapped to the weights. 
Finally, Re-SAT trains the models with the reweighted samples.

In summary, we observe the biased performance of ASR on disordered speech and analyze the ASR system through the lens of debiasing. Furthermore, we present sample affinity test (SAT) to directly measure the debiasing helpfulness of samples. We also propose Re-SAT, a novel method for sample reweighting that is available under the group label-free environment and mitigates the margin between the ASR performances of healthy speakers and dysarthric speakers.

\section{Related work} 
\textbf{Dysarthric speech recognition} ASR on dysarthric speech still remains a challenging problem. Since the disordered speech dataset has limited scalability due to the difficulty of data collection, data augmentation on dysarthric speech has been widely studied~\cite{geng2022investigation, bhat2022improved, jin2021adversarial, jin2022adversarial, vachhani2018data}. Other investigations adopt pre-trained self-supervised learning models \cite{baskar2022speaker, hu2023exploring, violeta2022investigating, hernandez2022cross} for improved disordered speech recognition. On the other hand, \cite{geng2022fly, geng2022spectro, geng2022speaker} utilize some prior knowledge that a dysarthric speaker has distinct characteristics compared to healthy or other dysarthric speakers by leveraging speaker adaptation using the spectro-temporal level. In this study, we approach ASR on dysarthric speech from a different point of view, debiasing, which has never been used to the best of our knowledge.

\textbf{Debiasing} Debiasing is a challenging but essential area for fair and inclusive deep learning. Group robustness, which denotes an ability to perform satisfactorily across different groups in terms of the given task, is one of the key factors of a debiased model. We focus on group robustness without group annotations, for a wider real-world application. Sample reweighting is one of the promising methods for debiasing. While empirical risk minimization (ERM) uniformly averages all of the performances across the given samples, sample reweighting methods such as Learning from Failure (LfF)~\cite{nam2020learning} and Just Train Twice (JTT)~\cite{liu2021just} upweight the samples that are expected to belong to the poor performance group called bias-conflicting samples. These methods are based on the intuition that upweighting the bias-conflicting samples with large losses leads to a debiased model, which is not always true.
In particular, we test ASR systems on dysarthric speakers, each of who can be regarded as a respective group due to the variety of dysarthria. 
In this complex real-world environment, reweighting the samples using a fine-grained metric is an important issue.
For this reason, we aim to propose direct and accurate criteria for reweighting beyond the estimation of bias-conflicting samples based on loss.

\textbf{Task affinity} Given a set of tasks in a multi-task learning setup, task grouping aims to find the clusters of tasks that should be trained simultaneously for improved performance. Task affinity~\cite{fifty2021efficiently} is proposed to address the task grouping problems. To examine the task affinity of task $i$ on task $j$, they leverage a lookahead update on task $i$. Then, they compute the loss shift after the lookahead update to explore the effect of task $i$ on $j$. Inspired by task affinity, we propose sample affinity to investigate the inter-sample effect for debiasing.

\begin{figure*}
    \centering
    \includegraphics[width=1\textwidth]{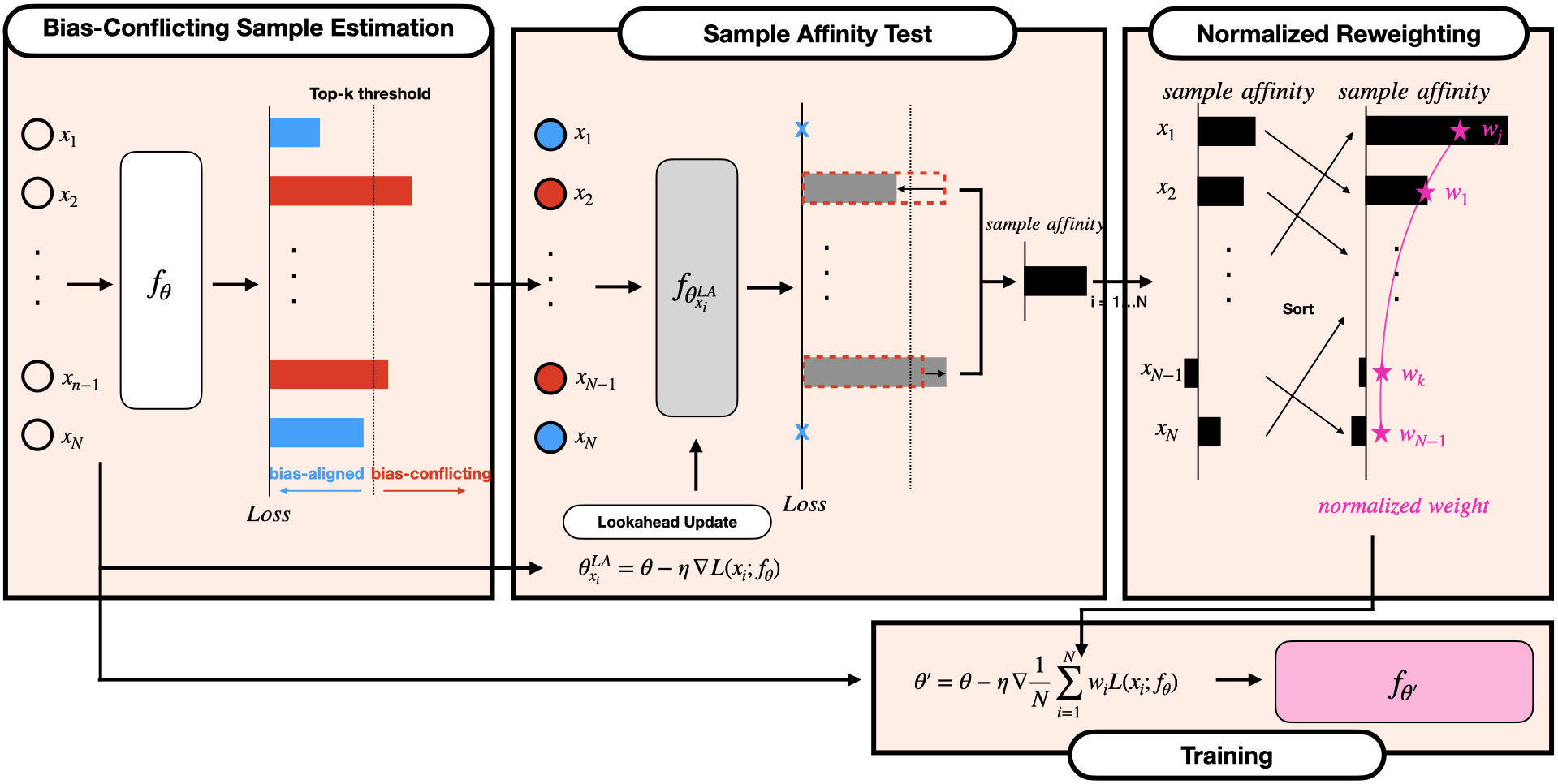}
    \caption{Illustration of Re-SAT for debiased dysarthric speech recognition. Re-SAT consists of 1) bias-conflicting sample estimation, 2) sample affinity test, 3) normalized reweighting, and 4) training blocks.}
    \label{fig:resat}
\end{figure*}

\section{Method}
In this section, we introduce our sample reweighting with sample affinity test (Re-SAT), which is proposed to challenge the ASR on dysarthric speech. Re-SAT consists of four components: 
1) bias-conflicting sample estimation, 
2) sample affinity test, 
3) normalized reweighting and 
4) training, as shown in Figure \ref{fig:resat}.
The data samples are reweighted through the first three components, and then Re-SAT trains the model with the reweighted data samples in the 4) training block.
In summary, Re-SAT upweights the bias-blocking samples and downweights the bias-accelerating samples using sample affinity test for debiasing. Details are as follows.

\subsection{Bias-conflicting sample estimation}
At the first step, Re-SAT estimates the bias-conflicting samples using the losses of samples in each batch. Re-SAT regards the samples with the largest $K$ losses in a batch as the bias-conflicting samples. Unlike the previous research~\cite{liu2021just} which fixes the estimation results, Re-SAT keeps estimating the bias-conflicting samples during training to accurately reflect the current state of the model. That is, the estimation result for each sample can be modified throughout training.

Although Re-SAT estimates the bias-conflicting samples, Re-SAT does not map the estimation results for reweighting because the large loss does not guarantee the debiasing ability, as we introduced in the above sections. We address this issue by designing a novel criterion that directly estimates the bias-blocking samples, for debiasing ability.

\subsection{Sample affinity test}
We propose a novel sample affinity test (SAT) to evaluate the debiasing ability of samples to filter the bias-accelerating samples.
Given the model $f_\theta$, the samples in the mini-batch $x_1,...,x_N$, and the estimated bias-conflicting samples $\hat b_1,..., \hat b_K$, SAT utilizes single step lookahead updating with respect to $x_i$ as 
\begin{equation}
    \theta^{LA}_{x_i}=\theta - \eta\nabla L(x_i;f_\theta),
\end{equation}
for $i=1,...,N$ where $\eta$ is learning rate, $N$ is batch size, and $L$ is a loss function. 
After the lookahead updating process, SAT compares the loss of 1) the lookahead model $\theta^{LA}_{x_i}$ and 2) the original model $\theta$ on the bias-conflicting samples to calculate the averaged sample affinity:
\begin{equation}
    \text{SA}(x_i\rightarrow \{\hat{b}_1, ..., \hat{b}_K\};f_\theta) = \frac{1}{K} \sum_{\forall k}\left(1-\frac{L(\hat b_k;f_{\theta^{LA}_{x_i}})}{L(\hat b_k;f_\theta)} \right).
\end{equation}

We present sample affinity inspired by task affinity~\cite{fifty2021efficiently} for debiasing. Note that SAT computes the sample affinity every step to take into account the current state of the model, while task affinity is averaged across the whole training step.
 
\subsection{Normalized reweighting}
Sample affinity on the bias-conflicting samples approximates the bias-blocking effect of a given sample. However, sample affinity is not an absolute but relative score which depends on the bias-conflicting sample set, the current state of the model, and the maturity of the model. Therefore, Re-SAT does not directly map the sample affinity to weights. Instead, Re-SAT normalizes the sample affinity by extracting the rank of the sample affinity with respect to the descending order within the batch. This normalization stabilizes unwanted shifts in the learning rate.

For sorted samples $x_1,...,x_r,...,x_N$ with respect to sample affinity in descending order, Re-SAT finally reweights the samples using the function $w: \mathbb{N}\rightarrow\mathbb{R}^+$, which is defined as follows:
\begin{equation}
    w(r)=\frac{\exp(s(N-r)/(N-1))}{\sum_{r=1}^N\exp(s(N-r)/(N-1))}
\end{equation}
for $r=1,...,N$ where $r$ is the rank of the sample $x_r$ and $s$ is the constant hyperparameter.
For our experiments, we set batch size $N=32$ and $s=4$.

\subsection{Training}
In this component, Re-SAT trains the model $f_\theta$ with the reweighted data samples. Unlike the vanilla training that averages the loss of each data sample, Re-SAT leverages weighted average for total loss as:
\begin{equation}
    \theta'=\theta-\eta\nabla\dfrac{1}{N}\sum_{r=1}^N w(r)L(x_r;f_\theta).
\end{equation}
for sorted samples $x_1, ..., x_N$ with respect to sample affinity $\text{SA}(x_i)$ in descending order.

\section{Experiments}
\subsection{Dataset}
UASpeech corpus \cite{kim2008dysarthric} is used for our experiments.
UASpeech corpus is one of the largest English dysarthric speech datasets, composed of speech from 15 dysarthric speakers and 13 healthy control speakers. Each dysarthric speaker is classified into very low, low, mid, and high levels of intelligibility. The dataset consists of three blocks B1, B2, and B3. Each block contains 155 common words and 100 uncommon words spoken by dysarthric and healthy control speakers. 
We used block 1 and block 3 of all healthy and dysarthric speech recorded on microphone M5 as a training set, and all data in block 2 as the test set. Note that we used not only dysarthric speech but also healthy speech as the test set. Basically, UASpeech provides the speech data denoised by \textit{noisereduce} \cite{sainburg2020finding, tim_sainburg_2019_3243139}; therefore we used this version for our experiments. We excluded the uncommon words from both the training and test set and no additional data augmentation was conducted.

\subsection{Model}
We fine-tuned \textit{Whisper} \cite{radford2022robust}, the recent state-of-the-art ASR model, trained on a large amount of labeled audio-transcription data. It employs a simple encoder-decoder Transformer architecture.
We used the \textit{Whisper-tiny}, pre-trained on English only with 39M parameters, which is available at HuggingFace transformers repository \cite{wolf2019huggingface}.

\subsection{Experimental Setups}
We trained the model using AdamW optimizer~\cite{loshchilov2018decoupled} with a learning rate of $10^{-5}$, weight decay of 0.1, and batch size of 32 for 30 epochs. For the proposed method, Re-SAT, we investigated the effect of the number of bias-conflicting samples in batch as $K\in\{2, 4, 8, 16\}$. For comparison, we tested the JTT method \cite{liu2021just} with upweighting value 25 and bias-conflicting sample identification epoch 3. For the ablation studies, we tested the model named Re-Loss, which utilizes the loss-based sample reweighting instead of sample affinity and follows Re-SAT in other settings, to figure out the impact of the sample affinity test.

\section{Results}
\label{results}

\begin{table}[t]
\centering
\begin{tabular}{ll|cccc}
\multicolumn{1}{l|}{} &  & \multicolumn{4}{c}{WER (\%)} \\ \hline
\multicolumn{1}{l|}{} & \begin{tabular}[c]{@{}l@{}}Speaker\\ (Intelligibility \%)\end{tabular} & ERM & JTT & Re-Loss & \begin{tabular}[c]{@{}c@{}}Re-SAT\\ (Ours)\end{tabular} \\ \hline
\multicolumn{1}{l|}{\multirow{5}{*}{VL}} & M04 (2\%) & 84.39 & \textbf{79.87} & 82.45 & 80.26 \\
\multicolumn{1}{l|}{} & F03 (6\%) & 44.33 & 42.21 & 40.28 & \textbf{38.25} \\
\multicolumn{1}{l|}{} & M12 (7\%) & 44.41 & 42.04 & 41.72 & \textbf{35.38} \\
\multicolumn{1}{l|}{} & M01 (17\%) & 50.97 & 51.77 & 47.10 & \textbf{44.19} \\ \cline{2-6} 
\multicolumn{1}{l|}{} & Avg. & 54.66 & 52.46 & 51.50 & \textbf{48.09} \\ \hline
\multicolumn{1}{l|}{\multirow{4}{*}{L}} & M07 (28\%) & 16.22 & 23.78 & 16.96 & \textbf{16.04} \\
\multicolumn{1}{l|}{} & F02 (29\%) & 19.72 & 22.86 & 16.59 & \textbf{15.94} \\
\multicolumn{1}{l|}{} & M16 (43\%) & 22.04 & 23.55 & 20.32 & \textbf{19.57} \\ \cline{2-6} 
\multicolumn{1}{l|}{} & Avg. & 19.19 & 23.39 & 17.84 & \textbf{17.06} \\ \hline
\multicolumn{1}{l|}{\multirow{4}{*}{M}} & M05 (58\%) & \textbf{12.63} & 13.55 & 12.90 & 12.72 \\
\multicolumn{1}{l|}{} & M11 (62\%) & 14.30 & 12.80 & 12.04 & \textbf{10.11} \\
\multicolumn{1}{l|}{} & F04 (62\%) & 13.88 & 13.88 & \textbf{13.22} & 14.64 \\ \cline{2-6} 
\multicolumn{1}{l|}{} & Avg. & 13.57 & 13.44 & 12.75 & \textbf{12.59} \\ \hline
\multicolumn{1}{l|}{\multirow{6}{*}{H}} & M09 (86\%) & 9.86 & 11.15 & \textbf{8.66} & 8.76 \\
\multicolumn{1}{l|}{} & M14 (90\%) & 12.17 & 13.73 & 11.98 & \textbf{9.86} \\
\multicolumn{1}{l|}{} & M10 (93\%) & 3.50 & 3.32 & \textbf{2.40} & 2.76 \\
\multicolumn{1}{l|}{} & M08 (95\%) & 4.61 & 5.53 & 4.70 & \textbf{3.78} \\
\multicolumn{1}{l|}{} & F05 (95\%) & \textbf{2.03} & 4.33 & \textbf{2.03} & 2.76 \\ \cline{2-6} 
\multicolumn{1}{l|}{} & Avg. & 6.43 & 7.61 & 5.95 & \textbf{5.59} \\ \hline
\multicolumn{2}{l|}{\begin{tabular}[c]{@{}l@{}}Avg.\\ (Dysarthric)\end{tabular}} & 21.49 & 22.25 & 20.15 & \textbf{19.93} \\ \hline
\multicolumn{2}{l|}{\begin{tabular}[c]{@{}l@{}}Avg.\\ (Healthy)\end{tabular}} & 3.81 & 4.21 & 3.75 & \textbf{3.40} \\ \hline
\multicolumn{2}{l|}{\begin{tabular}[c]{@{}l@{}}Avg.\\ (Overall)\end{tabular}} & 12.93 & 13.51 & 12.20 & \textbf{12.08}
\end{tabular}
\label{table:tiny}
\caption{ASR results in terms of Word Error Rate (WER) on UASpeech corpus. VL/L/M/H refer to very low/low/mid/high intelligibility groups, respectively.
The percentage of intelligibility is measured based on how accurately naive human listeners can transcribe isolated words produced by speakers, according to \cite{kim2008dysarthric}. $K$ of Re-SAT is set to 4.
}
\end{table}

\begin{table}[t]
\centering
{\footnotesize
\begin{tabular}{l|cccccc|c}
 & \multicolumn{6}{c|}{Intelligibility group} &  \\ \hline
K & VL & L & M & H & \multicolumn{1}{c|}{\begin{tabular}[c]{@{}c@{}}Avg.\\ (D)\end{tabular}} & \begin{tabular}[c]{@{}c@{}}Avg.\\ (H)\end{tabular} & Avg. \\ \hline
ERM & 54.66 & 19.19 & 13.57 & 6.43 & \multicolumn{1}{c|}{21.49} & 3.81 & 12.93 \\ \hline
2 & 50.12 & 17.49 & 14.15 & 5.75 & \multicolumn{1}{c|}{19.93} & 3.72 & 12.08 \\
4 & \textbf{48.09} & \textbf{17.06} & \textbf{12.59} & \textbf{5.59} & \multicolumn{1}{c|}{\textbf{19.05}} & \textbf{3.40} & \textbf{11.47} \\
8 & 49.44 & 17.61 & 12.62 & 6.36 & \multicolumn{1}{c|}{19.75} & 3.84 & 12.05 \\
16 & 55.31 & 21.10 & 13.60 & 6.91 & \multicolumn{1}{c|}{22.21} & 3.96 & 13.36
\end{tabular}
}
\label{table:ablation}
\caption{Comparison on hyperparameter $K$ with word error rate (WER). \text{D} and \text{H} refer to dysarthric and healthy speech, resepctively.}
\end{table}

We investigated with the effect of the proposed method, Re-SAT, by comparing it to empirical risk minimization (ERM) and JTT for debiasing. In contrast to most of the dysarthric speech recognition studies that focus only on dysarthric speech, we investigated the effect of Re-SAT to ASR system on healthy speech for a fair comparison. This is important for debiasing because the performance disparity on dysarthric speech and healthy speech determines the group robustness of the systems. 

Results in Table 1 show word error rate (WER) of the models on each speaker and intelligibility group. 
It is shown that Re-SAT contributes to performance improvements for each group compared to ERM by relatively decreasing 12.02\% (54.66 $\rightarrow$ 48.09), 11.10\% (19.19 $\rightarrow$ 17.06), 7.22\% (13.57 $\rightarrow$ 12.59), and 13.06\% (6.43 $\rightarrow$ 5.59) in terms of the averaged WER for the very low, low, mid, and high intelligibility group, respectively. Surprisingly, Re-SAT also surpasses ERM on the healthy speaker group.
From a speaker-wise view, it is observed that the performances of 12 out of 15 speakers show enhanced results in terms of WER on the Re-SAT-based ASR model. These results demonstrate the robust performance gain of Re-SAT even under a diverse group environment. On the contrary, the other debiasing method, JTT, increases the averaged WER on healthy speech, low and high intelligibility speech although JTT reduces the very low and mid intelligibility group's WER, compared to those of ERM.

\begin{figure}[t]
    \centering
    \includegraphics[width=0.45\textwidth]{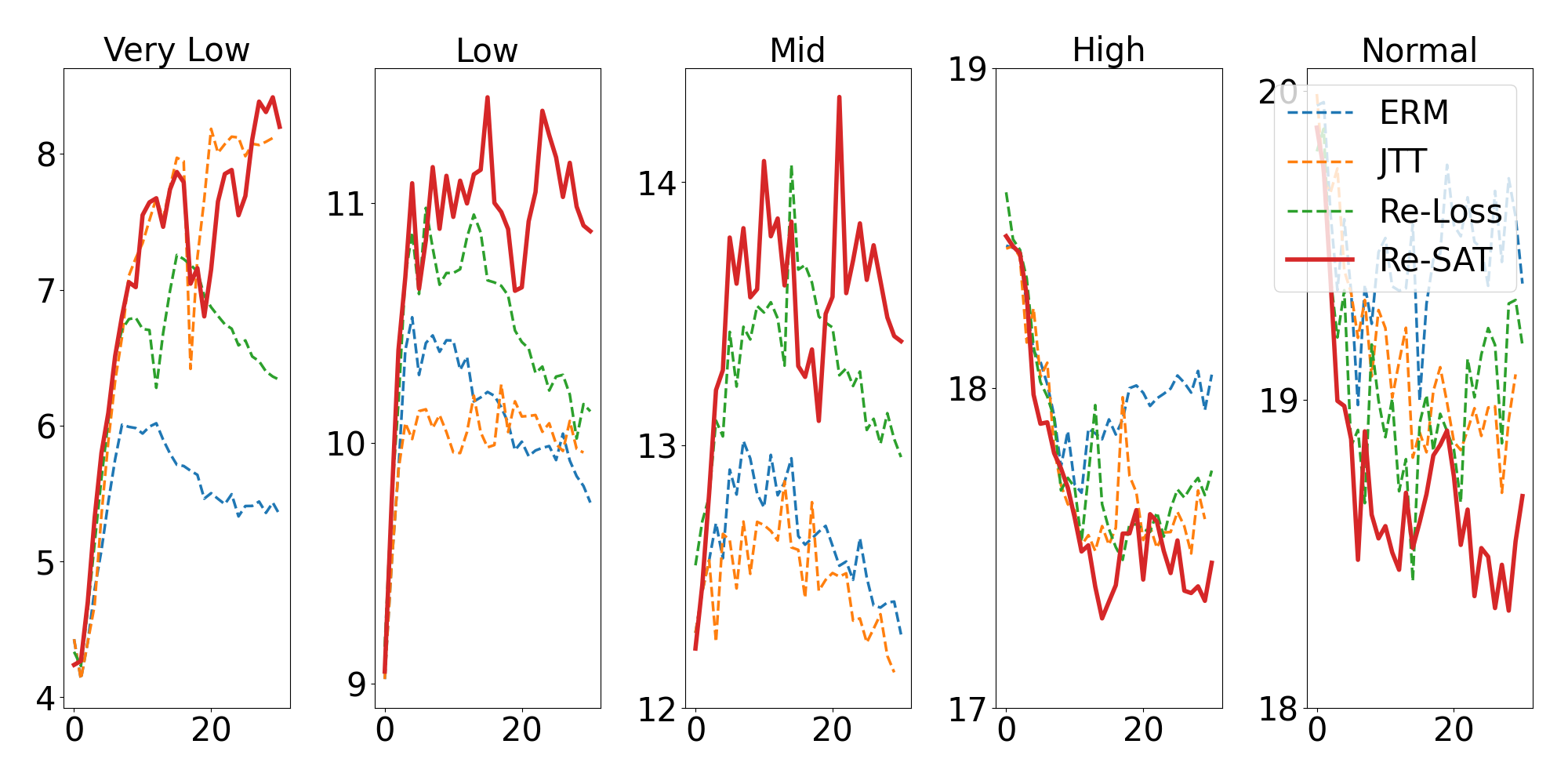}
    \caption{Illustrations of the averaged loss ranks  sorted by descending order in the mini-batch for each intelligibility group. The x-axis denotes training epochs and the y-axis denotes the loss rank in mini-batch.}
    \label{fig:rank}
\end{figure}

Figure \ref{fig:rank} shows the rank sorted in descending order of training loss within the batch with respect to each intelligibility group. For very low, low, and mid-intelligibility groups that are vulnerable to bias, the increasing and large rank value is desired because it means that the loss for the biased group is decreased, even though it is approximated by training loss. Interestingly, debiasing of JTT works only on the single worst (very low) intelligibility group and even deteriorates the low and mid groups' ranks. This problem can be caused by the structure of JTT's 1) binary classification of bias-conflicting samples and 2) which are fixed at an early stage. On the other hand, Re-SAT leverages 1) fine-grained reweighting and 2) updated bias-conflicting set through training and shows desirable results across diverse groups. Figure \ref{fig:rank} also demonstrates that ERM is inappropriate for debiasing. Although the ranks on very low, low, and mid intelligibility groups of ERM increase at the early stage for all training schemes, the ranks decrease at the late stage of ERM. Even worse, the healthy speaker's rank shows the highest rank compared to other methods, which means the biased performance of ERM. We also explore the Re-Loss, which substitutes the sample affinity-based reweighting of Re-SAT with loss-based reweighting, to investigate the importance of the sample affinity test. Although Re-Loss outperforms ERM and JTT by a large margin, Re-SAT still shows a surpassing performance. This demonstrates the precise bias-blocking sample estimation ability of Re-SAT, as we expected.

Intuitively, selecting decent bias-conflicting samples takes an important role in Re-SAT. For this reason, we investigate the size of the selected bias-conflicting sample set $K$ in Re-SAT as shown in Table 2. Re-SAT shows stable improvements across the intelligibility groups where $K=2, 4, 8$ compared to ERM for all intelligibility groups. This supports our design choice of Re-SAT which re-estimates the bias-conflicting samples through training. Even though Re-SAT selects small bias-conflicting samples, Re-SAT can re-estimate the new underfitted data sample as a bias-conflicting sample through training. On the contrary, Re-SAT with $K=16$ even degrades the ASR results, since setting a large $K$ as 16 interferes with the accurate estimation of the bias-conflicting samples.

\section{Conclusions}
In this study, we address a debiasing problem of ASR on dysarthric speech towards a fair and inclusive ASR system. In contrast to the previous research that focuses only on the ASR performance for dysarthric speakers, we explore the fairer validation system by analyzing the performance of healthy and dysarthric speech at the same time. To achieve our goal, we propose a novel debiasing method based on sample reweighting, Re-SAT. The ASR system using Re-SAT surpasses the other baselines across the diverse dysarthric speakers and shows a robust performance gain over various hyperparameters. For future work, it is worth exploring the integrated system of Re-SAT and other dysarthric speech recognition systems by leveraging the versatile structure of Re-SAT.


\bibliographystyle{IEEEtran}
\bibliography{mybib}

\end{document}